# Capacity of Block-Memoryless Channels with Causal Channel Side Information

Hamid Farmanbar and Amir K. Khandani

Coding and Signal Transmission Laboratory
Department of Electrical and Computer Engineering
University of Waterloo
Waterloo, Ontario, N2L 3G1
Email: {hamid,khandani}@cst.uwaterloo.ca

## Abstract

The capacity of a time-varying block-memoryless channel in which the transmitter and the receiver have access to (possibly different) noisy causal channel side information (CSI) is obtained. It is shown that the capacity formula obtained in this correspondence reduces to the capacity formula reported in [1] for the special case where the transmitter CSI is a deterministic function of the receiver CSI.

## I. INTRODUCTION

Motivated by the result reported in [1], this correspondence studies the capacity of a stationary and ergodic time-varying block-memoryless (BM) channel where the transmitter and the receiver have access to noisy causal channel side information (CSI).

The CSI at the transmitter (CSIT) and the CSI at the receiver (CSIR) can be different. The time variations of the channel are modeled as a set of channel states where the channel is at some state at each time instant. A time-varying (state-dependent) BM channel is memoryless between blocks, however, within each block the state and the channel conditioned on the state can have memory. For example, such a channel model applies to systems based on frequency hopping with slow mobility. This results in a quasi-static (block) fading channel model where the channel fading is static within a block and changes independently between blocks as the frequency hops to a different carrier. A formal definition of a state-dependent BM channel is given in Section II.

The capacity of time-varying BM channels with CSI at transmitter and receiver has been studied in [1] and the capacity is obtained for the case that the CSIT is a deterministic function of the CSIR. As an example, the scenario where the CSIT is a deterministic function of the CSIR occurs when the receiver quantizes its observation of the channel state and transmits it via a noiseless channel to the transmitter. However, when the feedback channel is noisy, the CSIT will no longer be a deterministic function of the CSIR. In this correspondence, we obtain the capacity for such a general case.

The key idea comes from the capacity results due to Shannon for state-dependent discrete memoryless channels with causal side information at the transmitter [2]. In the model considered by Shannon, the state of the channel is perfectly known at the transmitter and unknown at the receiver. Shannon's work was extended by Salehi [3] to the case that (possibly different) noisy versions of the CSI are available at the transmitter and at the receiver. It was later shown by Caire and Shamai [4] that the capacity with noisy CSI can be obtained from Shannon's original work by considering a new state-dependent channel with CSIT alphabet as the new state alphabet. It is worth mentioning that in our problem, since CSIT symbols are not available up to the end of the current block, applying Shannon's results [2] to super symbols corresponding to blocks would not yield the capacity.

We will use the following notations throughout the correspondence. Random vari-ables are denoted by upper case letters $(X)$ and their values are denoted by lower case letters $(x)$. The sequence of random variables $X_m, \ldots, X_n$ is denoted by $X_m^n$; and $x_m^n$ denotes a particular realization of $X_m^n$. The sequences $X_1^n$ and $x_1^n$ are denoted by $X^n$ and $x^n$, respectively. Sets are denoted by calligraphic letters $(\mathcal{X})$; $|\mathcal{X}|$ denotes the cardinality of $\mathcal{X}$, and $\mathcal{X}^n = \underbrace{\mathcal{X} \times \cdots \times \mathcal{X}}_{n}$ is the $n$-th Cartesian power of $\mathcal{X}$.

## II. CHANNEL MODEL

The channel model considered in this correspondence is the same as the one introduced in [1] where a state-dependent block-memoryless channel is defined by a finite channel input alphabet $\mathcal{X}$, a finite channel output alphabet $\mathcal{Y}$, a finite state alphabet $\mathcal{S}$, and transition probabilities $p(y^{n_0}|x^{n_0}, s^{n_0})$ where $n_0$ is the channel block length. We denote the CSIT and the CSIR by $U \in \mathcal{U}$ and $V \in \mathcal{V}$, respectively. The CSIT and the CSIR are dependent on the state according to the joint distribution $p(s^{n_0}, u^{n_0}, v^{n_0})$.

It is convenient to express the transition probabilities of the channel in terms of the CSIT and the CSIR as

$$\begin{aligned}
p(y^{n_0}, v^{n_0}|x^{n_0}, u^{n_0}) &= \sum_{s^{n_0}} p(y^{n_0}, v^{n_0}|x^{n_0}, u^{n_0}, s^{n_0}) p(s^{n_0}|x^{n_0}, u^{n_0}) \\
&= \sum_{s^{n_0}} p(y^{n_0}|x^{n_0}, u^{n_0}, s^{n_0}, v^{n_0}) p(v^{n_0}|x^{n_0}, u^{n_0}, s^{n_0}) p(s^{n_0}|x^{n_0}, u^{n_0}) \\
&= \sum_{s^{n_0}} p(y^{n_0}|x^{n_0}, s^{n_0}) p(v^{n_0}|u^{n_0}, s^{n_0}) p(s^{n_0}|u^{n_0}) \\
&= \sum_{s^{n_0}} p(y^{n_0}|x^{n_0}, s^{n_0}) p(s^{n_0}, u^{n_0}, v^{n_0})/p(u^{n_0}), \quad (1)
\end{aligned}$$

where $p(u^{n_0}) = \sum_{s^{n_0}, v^{n_0}} p(s^{n_0}, u^{n_0}, v^{n_0})$.

For $n = Jn_0$ uses of the channel, we have

$$p(y^n, v^n|x^n, u^n) = \prod_{j=0}^{J-1} p\left(y_{jn_0+1}^{(j+1)n_0}, v_{jn_0+1}^{(j+1)n_0} \big| x_{jn_0+1}^{(j+1)n_0}, u_{jn_0+1}^{(j+1)n_0}\right), \quad (2)$$

and

$$p(s^n, u^n, v^n) = \prod_{j=0}^{J-1} p\left(s_{jn_0+1}^{(j+1)n_0}, u_{jn_0+1}^{(j+1)n_0}, v_{jn_0+1}^{(j+1)n_0}\right). \quad (3)$$

We define a $(2^{nR}, n)$ block code of length $n$ for the state-dependent BM channel to be $2^{nR}$ sequences of $n$ encoding functions $f_i : \mathcal{W} \times \mathcal{U}^n \to \mathcal{X}$ for $i = 1, \ldots, n$ such that $x_i = f_i(w, u_1^i)$, where $w \in \mathcal{W} = \{1, \ldots, 2^{nR}\}$. Note that the channel input at time $i$ depends on the CSIT up to time $i$. In other words, we consider causal knowledge setting. At the receiver, a decoding function $g : \mathcal{Y}^n \times \mathcal{V}^n \to \mathcal{W}$ is used to decode the transmitted message as $\hat{w} = g(y_1^n, v_1^n)$. The rate of the block code is $R = \frac{1}{n} \log |\mathcal{W}|$, and $P_e^{(n)}$ is defined as the probability that a message $W$, uniformly distributed over $\mathcal{W}$, is received in error, i.e.,

$$P_e^{(n)} = \Pr\{\hat{W} \neq W\}. \tag{4}$$

## III. CAPACITY OF BLOCK-MEMORYLESS CHANNELS WITH CSI

The capacity of a time-varying BM channel for the case that the CSIT, $U^{n_0}$, is a deterministic function of the CSIR, $V^{n_0}$, is given by [1]

$$\begin{aligned} C &= \max_{p(x^{n_0}|u^{n_0})} \frac{1}{n_0} I(X^{n_0}; Y^{n_0} | V^{n_0}) \\ &= \sum_{u^{n_0}} p(u^{n_0}) \max_{p(x^{n_0}|u^{n_0})} \frac{1}{n_0} I(X^{n_0}; Y^{n_0} | u^{n_0}, V^{n_0}) \end{aligned} \tag{5}$$

where the maximum is taken over all distributions satisfying the causal side information constraint, i.e.,

$$p(x^{n_0}|u^{n_0}) = \prod_{i=1}^{n_0} p(x_i | x^{i-1}, u^i). \tag{6}$$

The capacity is achieved by a scheme that adapts itself to channel variations so that for every realization of the CSIT, the encoder uses a code which is capacity-achieving for that specific realization. The final coding scheme will be simply a multiplexed version of the coding schemes for all possible CSIT realizations.

The scenario in which the CSIT is a function of the CSIR describes a situation where the CSIT is, for example, a quantized version of the CSIR due to rate restrictions on the capacity of the feedback link between the receiver and the transmitter. However, when the feedback channel introduces noise, the CSIT will no longer be a deterministic function of the CSIR. In this case, the decoder will no longer know the transmission strategy and this complicates capacity analysis. In the following, we show that the Shannon's approach for state-dependent discrete memoryless channels with causal side information at the transmitter, with some modifications, can be used to obtain the capacity in this more general case. It should be notes that applying Shannon's scheme to our channel with super symbols of size $n_0$ does not yield the capacity since CSIT is available only up to the current symbol, not up to the end of the current channel block (super symbol). We will show that to achieve the capacity, it is sufficient to consider encoding schemes that use the CSIT up to the current symbol and within the current super symbol. In other words, there is no loss in capacity by disregarding the past CSIT symbols that are not within the current super symbol.

*Theorem 1:* The capacity of a time-varying BM channel with the CSIT and the CSIR denoted by $U^{n_0}$ and $V^{n_0}$, respectively, is equal to

$$C = \max_{p(t^{n_0})} \frac{1}{n_0} I(T^{n_0}; Y^{n_0} | V^{n_0}), \tag{7}$$

where the *equivalent* channel from $T^{n_0}$ to $(Y^{n_0}, V^{n_0})$ is defined by[1]

$$p(y^{n_0}, v^{n_0} | t^{n_0}) = \sum_{u^{n_0}} p(u^{n_0}) p\left(y^{n_0}, v^{n_0} | x_i = t_i(u^i)|_{i=1}^{n_0}, u^{n_0}\right). \tag{8}$$

*Proof:*

*Achievability*: Consider the following encoding scheme. A message $w \in \{1, \ldots, 2^{nR}\}$ is encoded to $\left(t_1^{n_0}(w), t_{n_0+1}^{2n_0}(w), \ldots, t_{(J-1)n_0+1}^{n}(w)\right)$, where $t_{jn_0+i} \in \mathcal{X}^{|\mathcal{U}|^i}$ is a function from $\mathcal{U}^i$ to $\mathcal{X}$ [2], $j = 0, 1, \ldots, J-1$, $i = 1, 2, \ldots, n_0$. Then, for any CSIT sequence $u_1^n$, the channel input sequence $x_1^n$ is given by $x_{jn_0+i} = t_{jn_0+i}\left(u_{jn_0+1}^{jn_0+i}\right)$, $j = 0, 1, \ldots, J-1$, $i = 1, 2, \ldots, n_0$. The new channel from $T^{n_0}$ to $(Y^{n_0}, V^{n_0})$ defined by (8) is not state dependent and for which the rate $\frac{1}{n_0}I(T^{n_0};Y^{n_0},V^{n_0})$ is achievable for a fixed $p(t^{n_0})$. However, we have

$$
\begin{aligned}
I(T^{n_0};Y^{n_0},V^{n_0}) &= I(T^{n_0};V^{n_0}) + I(T^{n_0};Y^{n_0}|V^{n_0}) \\
&= I(T^{n_0};Y^{n_0}|V^{n_0}),
\end{aligned} \tag{9}
$$

since $T^{n_0}$ is independent of $V^{n_0}$. Hence, the rate $C$ given in (7) is achievable.

---

[1]Theorem 1 may equaivalently be stated as follows. The capacity is given by (7) in which the maximization is restricted to distributions satisfying $p(t^{n_0}, u^{n_0}, v^{n_0}, x^{n_0}, y^{n_0}) = p(t^{n_0})p(u^{n_0})p(x^{n_0}|t^{n_0}, u^{n_0})p(y^{n_0}, v^{n_0}|x^{n_0}, u^{n_0})$ and $x_i = t_i(u^i)$, $i = 1, \ldots, n_0$. I.e., $T^{n_0}$ is independent of $U^{n_0}$ and $p(x^{n_0}|t^{n_0}, u^{n_0})$ takes values zero and one only.

[2]There is a one-to-one correspondence between the elements of $\mathcal{U}^i$ and the elements of $\{1, 2, \ldots, |\mathcal{U}|^i\}$. A function from $\mathcal{U}^i$ to $\mathcal{X}$ can be represented by a $|\mathcal{U}|^i$-tuple composed of elements of $\mathcal{X}$. Each component of the $|\mathcal{U}|^i$-tuple represents the value of the function for a specific element of $\mathcal{U}^i$.

*Converse*: For any $(2^{nR},n)$ code for the state-dependent BM channel with arbitrary small probability of error, we have

$$
\begin{aligned}
nR &= H(W) &(10)\\
&= I(W;Y^n,V^n) + H(W|Y^n,V^n) &(11)\\
&\leq I(W;Y^n,V^n) + n\epsilon_n &(12)\\
&= \sum_{j=0}^{J-1} I\left(W; Y_{jn_0+1}^{(j+1)n_0}, V_{jn_0+1}^{(j+1)n_0} | Y_1^{jn_0}, V_1^{jn_0}\right) + n\epsilon_n &(13)\\
&\leq \sum_{j=0}^{J-1} I\left(W, Y_1^{jn_0}, V_1^{jn_0}; Y_{jn_0+1}^{(j+1)n_0}, V_{jn_0+1}^{(j+1)n_0}\right) + n\epsilon_n &(14)\\
&\leq \sum_{j=0}^{J-1} I\left(W, U_1^{jn_0}; Y_{jn_0+1}^{(j+1)n_0}, V_{jn_0+1}^{(j+1)n_0}\right) + n\epsilon_n &(15)\\
&= \sum_{j=0}^{J-1} I\left(W, U_1^{jn_0}; Y_{jn_0+1}^{(j+1)n_0} | V_{jn_0+1}^{(j+1)n_0}\right) + n\epsilon_n &(16)\\
&= \sum_{j=0}^{J-1} I\left(T_j; Y_{jn_0+1}^{(j+1)n_0} | V_{jn_0+1}^{(j+1)n_0}\right) + n\epsilon_n &(17)\\
&\leq nC + n\epsilon_n, &(18)
\end{aligned}
$$

where $\epsilon_n = \frac{1}{n} + P_e^{(n)} R \to 0$ for large $n$; (12) follows from Fano's inequality; (15) follows from the data processing inequality for the Markov chain $(W, Y_1^{jn_0}, V_1^{jn_0}) \to (W, U_1^{jn_0}) \to (Y_{jn_0+1}^{(j+1)n_0}, V_{jn_0+1}^{(j+1)n_0})$; (16) follows since $(W, U_1^{jn_0})$ is independent of $V_{jn_0+1}^{(j+1)n_0}$; $T_j = (W, U_1^{jn_0})$; and (18) follows by comparing $I\left(T_j; Y_{jn_0+1}^{(j+1)n_0} | V_{jn_0+1}^{(j+1)n_0}\right)$ with (7) and noting that $T_j$ is independent of $U_{jn_0+1}^{(j+1)n_0}$ and $X_{jn_0+i} = f_{jn_0+i}\left(T_j, U_{jn_0+1}^{jn_0+i}\right)$, for $j = 0, \ldots, J-1$, $i = 1, \ldots, n_0$. ∎

In the sequel, we show that the capacity formula (7), reduces to (5) when $U^{n_0}$ is a deterministic function of $V^{n_0}$, i.e., $U^{n_0} = k(V^{n_0})$. Any distribution $p(t^{n_0})$ induces a distribution $p(x^{n_0}|u^{n_0})$ according to

$$p(x^{n_0}|u^{n_0}) = \sum_{t^{n_0}:t^{n_0}(u^{n_0})=x^{n_0}} p(t^{n_0}), \qquad \forall x^{n_0} \in \mathcal{X}^{n_0}, \forall u^{n_0} \in \mathcal{U}^{n_0}, \qquad (19)$$

where $t^{n_0}(u^{n_0}) = x^{n_0}$ implies $x_i = t_i(u^i)$, $i = 1, \ldots, n_0$. On the other hand, for any distribution $p(x^{n_0}|u^{n_0})$, there is a corresponding distribution $p(t^{n_0})$ which can be obtained by solving (19). Given a realization of the CSIT, $u^{n_0}$, we have the Markov chain $T^{n_0} \to X^{n_0}|u^{n_0} \to (Y^{n_0}, V^{n_0})|u^{n_0}$. Therefore, by averaging over all realizations, we have

$$I(T^{n_0}; Y^{n_0}, V^{n_0}|U^{n_0}) \leq I(X^{n_0}; Y^{n_0}, V^{n_0}|U^{n_0}). \qquad (20)$$

However,

$$\begin{aligned} I(T^{n_0}; Y^{n_0}V^{n_0}|U^{n_0}) &= I(T^{n_0}; V^{n_0}|U^{n_0}) + I(T^{n_0}; Y^{n_0}|V^{n_0}, U^{n_0}) \\ &= I(T^{n_0}; Y^{n_0}|V^{n_0}), \end{aligned} \qquad (21)$$

since $T^{n_0}$ is independent of $(U^{n_0}, V^{n_0})$, and $U^{n_0} = k(V^{n_0})$. Furthermore,

$$\begin{aligned} I(X^{n_0}; Y^{n_0}, V^{n_0}|U^{n_0}) &= I(X^{n_0}; V^{n_0}|U^{n_0}) + I(X^{n_0}; Y^{n_0}|V^{n_0}, U^{n_0}) \\ &= I(X^{n_0}; Y^{n_0}|V^{n_0}), \end{aligned} \qquad (22)$$

Since $V^{n_0} \to U^{n_0} \to X^{n_0}$ form a Markov chain. Hence,

$$\max_{p(t^{n_0})} I(T^{n_0}; Y^{n_0}|V^{n_0}) \leq \max_{p(x^{n_0}|u^{n_0})} I(X^{n_0}; Y^{n_0}|V^{n_0}). \qquad (23)$$

On the other hand,

$$I(T^{n_0}; Y^{n_0}|V^{n_0}) = H(Y^{n_0}|V^{n_0}) - H(Y^{n_0}|T^{n_0}, V^{n_0}) \qquad (24)$$
$$= H(Y^{n_0}|V^{n_0}) - H(Y^{n_0}|T^{n_0}, V^{n_0}, U^{n_0}) \qquad (25)$$
$$= H(Y^{n_0}|V^{n_0}) - H(Y^{n_0}|T^{n_0}, V^{n_0}, U^{n_0}, X^{n_0}) \qquad (26)$$
$$\geq H(Y^{n_0}|V^{n_0}) - H(Y^{n_0}|V^{n_0}, U^{n_0}, X^{n_0}) \qquad (27)$$
$$= H(Y^{n_0}|V^{n_0}) - H(Y^{n_0}|X^{n_0}, V^{n_0}) \qquad (28)$$
$$= I(X^{n_0}; Y^{n_0}|V^{n_0}), \qquad (29)$$

where (25) and (28) follow since $U^{n_0} = k(V^{n_0})$; (26) follows since $X^{n_0}$ is a function of $T^{n_0}$ and $U^{n_0}$; and (27) follows since conditioning reduces entropy. Hence,

$$\max_{p(t^{n_0})} I(T^{n_0}; Y^{n_0}|V^{n_0}) \geq \max_{p(x^{n_0}|u^{n_0})} I(X^{n_0}; Y^{n_0}|V^{n_0}). \qquad (30)$$

Comparing (23) and (30), we conclude the result.

## IV. CONCLUSION

In this work, we obtained the capacity of time-varying block-memoryless channels where (possibly different) noisy causal CSI is available at the transmitter and at the receiver. We showed that for the case that the CSIT is a deterministic function of the CSIR, the obtained result reduces to the capacity expression reported in [1].